\documentclass[fleqn,usenatbib]{mnras}

\usepackage{newtxtext,newtxmath,verbatim}

\usepackage[T1]{fontenc}

\DeclareRobustCommand{\VAN}[3]{#2}
\let\VANthebibliography\thebibliography
\def\thebibliography{\DeclareRobustCommand{\VAN}[3]{##3}\VANthebibliography}

\usepackage{graphicx}	
\usepackage{amsmath}	
\usepackage{amssymb}	
\usepackage{ulem}
\usepackage{xcolor}

\title[Fast Radio Bursts]{Sudden discharge of  young charged magnetars as a new model for FRBs}

\author[M. G. B. de Avellar]{
M. G. B. de Avellar,$^{1,3}$\thanks{E-mail: mgb.avellar@unifesp.br}
P. P. B. Beaklini,$^{2}$
S. P. Nunes,$^{3}$
P. H. R. S. Moraes,$^{3,4}$
{M. Malheiro}$^{3}$
\\
$^{1}$Departamento de F\'{\i}sica, Instituto de Ci\^encias Ambientais, Qu\'{\i}micas e Farmac\^euticas (ICAQF), Universidade Federal de S\~ao Paulo - UNIFESP, \\Rua S\~ao Nicolau no.210, Centro, 09913-030, Diadema - SP,Brazil\\
$^{2}$National Radio Astronomy Observatory, 1003 Lopezville Road, Socorro, NM 87801, United States of America \\
$^{3}$Instituto Tecnol\'ogico de Aeron\'autica, Pra\c ca Marechal Eduardo Gomes, 50 - Vila das Ac\'acias, S\~ao Jos\'e dos Campos - SP, 12228-900, Brazil \\
$^{4}$Universidade de S\~ao Paulo, Instituto de Astronomia, Geof\'isica e Ci\^encias Atmosf\'ericas, Rua do Mat\~ao 1226, 05508-090 S\~ao Paulo, SP, Brazil\\
}

\date{Accepted XXX. Received YYY; in original form ZZZ}

\pubyear{2015}

\begin{document}
\label{firstpage}
\pagerange{\pageref{firstpage}--\pageref{lastpage}}
\maketitle

\begin{abstract}

We propose a new model for Fast Radio Bursts (FRBs) based on a sudden discharge of a charged young magnetar, caused by the short falling timescale of oppositely charged particles onto the magnetar. In this scenario, curvature radiation is emitted by particles accelerated at relativistic by the strong electric fields produced by the disconnection and the subsequent reconnection of the magnetic field lines, a process triggered by the sudden discharge. We modeled the magnetars as charged neutron stars in the static approximation using the exterior metric by the Reissner-Nordstr\"om. We also adopted an electrical charge distribution proportional to the mass-energy density, although our results are not strongly sensitive to the specific star’s charge distribution, only to the total charge. Our calculations show that the discharge and emission timescales are several milliseconds, compatible with the FRB phenomena for magnetars with a total charge of $\sim 10^{20}~\mathrm{C}$ and mass and radius in the range of 1.5 to 3.0 $\mathrm{M_{\odot}}$ and 10 to 45 km, respectively. Furthermore, the calculated total emitted power of a coherent pulse is $P_{tot}\sim 10^{42-43} erg/s$, and the frequency range and time scale are also consistent with FRBs astronomical observations. Finally, if the magnetar does not collapse after the discharge, the existence of FRB repeater sources can not rule out the existence of a blast repetition after the time needed to magnetosphere recharges and produce a new discharge.

\end{abstract}




\section{Introduction}
\label{sec:intro}



Fast Radio Bursts (FRBs) are mysterious astrophysical phenomena with unknown origins, first discovered in 2007 \citep{lorimer01} and whose emission has a set of distinctive characteristics such as it is fast with a duration typically between 1 and 10 milliseconds, bright in radio, 50 mJy to 100 Jy, and detected in the frequency range between 400 MHz and 8 GHz. Furthermore, the Dispersion Measure (DM) of these pulses, higher than the maximum value expected for the Milky Way, implies an extragalactic origin with distances that could be as far as 6 Gpc (z = 0.96) \citep[see, for example,][]{thornton01}. However, that set of characteristics defining an FRB is not very strict in the sense that a source can be classified as Rotating Radio Transient (RRAT) if the DM falls in the range expected for our galaxy \citep{mcLaughlin01}.

Until 2018, the population of known FRBs was about 60 sources \citep{petroff01}, with an estimated occurrence rate of $\sim 10^{3}$ detectable FRBs per day in the whole sky. Now, with the advent of CHIME - Canadian Hydrogen Intensity Mapping Experiment Fast Radio Burst - and the CHIME/FRB Collaboration \citep{CHIME01, CHIME02}, the number of FRBs increased one order of magnitude only in the first year of the experiment; now, there are 536 FRB sources between 400 and 800 MHz from 2018 July 25 to 2019 July 1 including the previous 60 sources \citep{2021_CHIME03}. These numbers, together with the estimated rate of about 800 /sky/day above a fluence of 5 Jy ms at 600 MHz \citep{2021_CHIME03} and 1700 /sky/day above a fluence of 2 Jy ms at 1.4 GHz \citep{2018_Bhandari01}, suggest that their progenitor must be prevalent in the Universe.

We also expect several detections with the BINGO radio telescope, which should start operation in 2023 \citep{2021_AbdallaE}. Its expected detection rate is one event with SNR $>3$ every 109 hours. Thus, despite its estimated detection rates scores lower than interferometers, which expect a few events a day, it scores better than most single dish instruments and has the advantage of being a transit telescope, covering about 5300 square degrees every day.

Although prevalent, the FRB nature is still elusive on theoretical grounds, so elusive that Platts et al. published ``A Living Theory Catalogue for Fast Radio Bursts'' \citep{platts01}. The authors describe the progenitor theories published so far, dividing them into categories. The categories are compact object mergers/interactions, the collapse of compact objects, supernovae remnants, active galaxy nuclei, collision and close encounters, and other models. To make things even more complicated, most of the time, FRBs are one-time events: to date, we observed only two repeaters \citep{spitler01, CHIME01, CHIME02}. 

The mechanism originating the FRBs discussed in our work is in the first category, compact object mergers/interactions, which includes neutron star–neutron star mergers/interactions, neutron star–supernova interactions, neutron star–white dwarf mergers, binary white dwarf merger, white dwarf–black hole mergers, neutron star–black hole mergers, pulsar–black hole interactions, and Kerr–Newman–black hole interactions. \citep[See][and references therein for the stimulating progenitor theories so far elaborated.]{platts01} 

We present a new theory for FRBs that keeps some models' explanatory power and main features while more naturally explaining the dichotomy repeater/non-repeater. Our model consists of a sudden discharge of a charged neutron star instead of a charged black hole, as in \cite{liuT01}, placing our theory closer to recent findings. (The progenitor theory of \cite{liuT01} is in the category Kerr-Newman-black hole interaction.) 

One of the most promising theories of FRBs relates these objects to magnetars \citep{CHIME02, 2021_Nicastro01, 2020_ZhangB, 2021_ZhangR, 2021_LiC}. In short, magnetars are young neutron stars whose magnetic fields reach $10^{13}$ G to $10^{15}$ G. Their magnetic field's decay leads them to produce enormous bursts and flares of X-rays and gamma-rays from time to time. We claim that some young magnetars are charged neutron stars that meet the proper conditions to sustain a magnetosphere induced to collapse. 


Thus, in Section 2, we describe the mathematical theory of the spacetime and particle orbits around a charged neutron star and the system's initial state that leads to a discharge of the neutron star generating the FRB. We also discuss the theory of charged neutron stars, following \cite{ray01, 2004_RayS}, to substantiate our claim that FRBs are charged magnetars, i.e., charged neutron stars that suddenly discharge.

We detail our method in Section 3, where we demonstrate that the falling times of oppositely charged particles on the magnetar are of the order of milliseconds, short enough to trigger a shockwave through the plasma producing the radio burst if the environment meets appropriate conditions. 

We discuss how our theory fits the observations in Section 4, mentioning recent observations concerning the Soft-Gamma Repeater SGR 1935+2154, the first event where a soft gamma burst occurred almost concomitantly with a radio counterpart.

We conclude by reinforcing that our model can explain the main properties of FRBs for a range of parameters of young charged neutron stars, for example, with masses and radii M $\simeq$ 2.5 $\mathrm{M_{\odot}}$ and R = [12 - 30] km, and give some prospects for the future.

\section{The system set-up}
\label{theSystem}

The keystone of the proposed mechanism is the falling of the oppositely charged particles onto the neutron star whose timescale is short enough to trigger a shock wave that excites the plasma, producing the burst in a milliseconds timescale. For this to happen, charges must be in the region of unstable orbits around the neutron star which is the region between the surface of the neutron star at radius R and $r_{m}$ (satisfying $r_{m}>R$, of course) given by $dV/dr = 0$, where $V(r)$ is the effective potential at the equatorial plane ($\theta=\pi/2$). $V(r)$ is given by the minimum allowed value of the energy E in the energy equation, which governs the particles' radial motion.

Due to the slow rotation in our magnetar model, we can describe the space outside the charged neutron star with the Reissner-Nordstr\"om solution. The line element employed for our calculations is of the form below \citep[see][]{misner01}:

\begin{equation}
    ds^{2} = -\frac{\Delta}{r^{2}}dt^{2} + sin^{2}(\theta)r^{2}~d\phi^{2}+\frac{r^{2}}{\Delta}dr^{2}+r^{2}d\theta^{2},
\end{equation} where

\begin{equation}
    \Delta = r^{2}-2Mr+Q^{2},
\end{equation}
being $M$ the mass, and $Q$ is the dimensionless charge of the neutron star. It is worth mentioning that as for the black hole case, the $M^{2}\geqslant Q^{2}$ relation must hold. 

The energy equation and the effective potential, $V(r)$, read \citep[see][with $a\rightarrow 0$]{liuT01}:

\begin{equation}
    E = \frac{\beta}{\alpha} + \frac{\sqrt{\beta^{2}-\alpha\gamma_{0}+\alpha r^{4}(p^{r})^{2}}}{\alpha},
\end{equation}

\begin{equation}
    V(r) = \frac{\beta}{\alpha} + \frac{\sqrt{\beta^{2}-\alpha\gamma_{0}}}{\alpha},
\end{equation} where 

\begin{equation}
    \alpha = r^{4} > 0,
\end{equation}

\begin{equation}
    \beta = e~Q~r^{3},
\end{equation}

\begin{equation}
    \gamma_{0} = (e~Q~r)^{2}-L_{z}^{2}\Delta - m^{2}r^{2}\Delta,
\end{equation}
being $m$, $e$, $L_z$ and $p^{r}$ the particle rest mass, charge, axial component of the angular momentum and radial components of the four-momentum, respectively. In \cite{liuT01}, the authors used as a benchmark the values $m=10^{-15}$ and $e=L_{z}=10^{-10}$ which we are going to follow.

On the equatorial plane ($\theta=\pi/2$), the time, $t$, and radial, $r$, components of the four-momentum of test particles can be written as \citep[see][]{misner01}:

\begin{equation}
    r^{2}p^{t} = \frac{r^{2}}{\Delta}P,
\end{equation}

\begin{equation}\label{p_r}
    r^{2}p^{r} = \sqrt{P^{2}-\Delta[m^{2}r^{2}+L_{z}^{2}]},
\end{equation} 
where $P$ is a function defined as
\begin{equation}
    P = r^{2}E-e~Q~r.
\end{equation}

The solution of $p^{r}=0$ at the chosen falling radius $r_{o}$ gives the minimum energy $E_{min}$ for the orbit from where the test particle falls onto the neutron star.

With values for $E \geq E_{min}$ and once we find $r_{m}$ (given by $dV/dr = 0$), we can calculate the falling time by

\begin{equation}
\label{deltaT}
    \delta t = - \int^{R}_{r_{o}}\frac{p^{t}}{p^{r}}dr,
\end{equation} where $R < r_{o} < r_{m}$, and the conversion to physical units is given by the factor $t_{cgs}=GM/c^{3}$ [sec].

Due to our magnetar scenario's slow rotation periods for the neutron stars, we described the exterior spacetime around a neutron star by the Reissner-Nordstr\"om metric. It is known that the multipolar structure of rotating black holes and neutron stars can significantly affect the observation of the electromagnetic and gravitational radiation from these objects \citep{paschalidis01,berti01}. However, we associate these objects to young magnetar pulsars with rotation periods greater than 0.02 seconds, which means spin frequencies smaller than 50 Hz, justifying our metric choice.

\subsection{Charged neutron stars}

In light of recent findings of an FRB in our galaxy \citep{petroff02}, we suggest that FRBs could be the observational evidence of the existence of young magnetars with spins roughly between 16 and 50 Hz as charged neutron stars that suddenly discharge or lose their charge partially or totally. \cite{ray01} give the theoretical foundation of charged neutron stars, and up to now, no observation constraint has been set. Thus, in the following sections, we intend to demonstrate that magnetar discharge could lead to a radiation emission on the same patterns as detected in the FRBs. 

\begin{figure}
\includegraphics[scale=0.37]{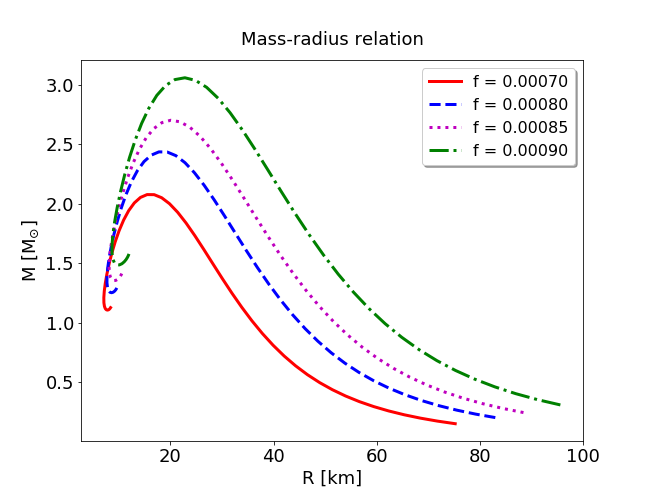}
\caption{Mass-radius relation.}
\label{fig:MR}
\end{figure}

Once the exterior solution for the FRBs is well defined, we specify the interior solution of the neutron star. As it is already known, the Birkhoff theorem hampers the interior of a rotating and charged neutron star to be described by Kerr-Newman solutions. Since magnetars have very long periods in the neutron stars period scale, validating the static approximation, the Reissner-Nordstr\"om metric is justified for also the interior, whose solutions follow from \cite{bekenstein}

\begin{align}\label{dq}
\frac{dQ(r)}{dr}=& \;4\pi\rho_{ch}(r) r^{3} \Delta^{-1/2},\\
\frac{dM(r)}{dr}=&\; 4\pi\rho(r) r^2  +\frac{Q(r)}{r}\frac{dQ(r)}{dr},\label{dm}\\
\frac{dp(r)}{dr}=&-(p(r)+\rho(r))\left[4\pi rp(r)+\frac{M(r)}{r^2}-\frac{Q(r)^2}{r^3}\right]\frac{r^2}{\Delta}\\\nonumber
&+\frac{Q(r)}{4\pi r^4}\frac{dQ(r)}{dr},\label{dp}
\end{align}
where $\rho_{ch}$ is the charge density, $p$ is the local pressure, and $\rho$ the energy density. The structure of the neutron star is obtained by integrating Eqs. (\ref{dq})-(\ref{dp}) using a fourth-order Runge-Kutta method. We assume an equation of state (EoS) for the stellar fluid obeying the relation
\begin{align}
p=K\rho^{1+1/n},
\end{align}
being $K$ and $n$ the polytropic constant and index.  Several papers have studied this kind of polytropic EoS \citep[see][and references therein]{Arbanil_2013, Arbanil_2018} and \cite{ray01} have used the polytrope $n = 3/2$ to describe the stellar fluid. However, it is worth mentioning that in charged compact objects theories (e.g., \cite{ray01,Negreiros2009Oct}), the charge influences the energy-momentum tensor, but it is neglected in the EoS. Here, we assume only that the charge density is proportional to the energy density as in \cite{ray01,Arbanil_2013,Arbanil_2018} 
\begin{align}
    \rho_{ch}(r)=f \rho(r),
\end{align}
being $f$ the charge fraction. The chosen charge profile mostly does not change the magnitude of the total charge of the stars, i.e., employing a different charge profile would yield the same order of the total charge for the charged stellar system, $Q\sim 10^{20}$C, a value which already can affect the compact star mass and radius, see Figures \ref{fig:MR} and \ref{fig:QM} and \citet{Negreiros2009Oct,Arbanil2015Oct}, for example.

Therefore, summarizing the strategy so far, we solved a modified TOV\footnote{TOV stands for Tolman-Oppenheimer-Volkoff equations, and these are the general relativistic version of the hydrostatic equilibrium. The modification in the original TOV comes from the Maxwell-Einstein stress tensor.} for a polytropic equation of state of index $n=3/2$ under the assumption that the charge density is proportional to matter density, i.e., $\rho_{ch}\sim \pm \rho_{m}$. The modified TOV was obtained in the framework of the Reissner-Nordstr\"om metric. In this way, there are no discontinuity or matching condition problems between the inner and outer metrics.

\begin{figure}
\includegraphics[scale=0.37]{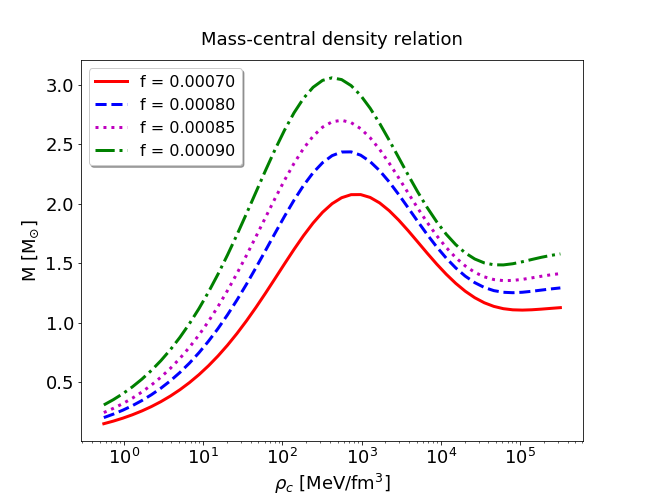}
\caption{Mass-central density relation.}
\label{fig:MRho}
\end{figure}

\begin{figure}
\includegraphics[scale=0.37]{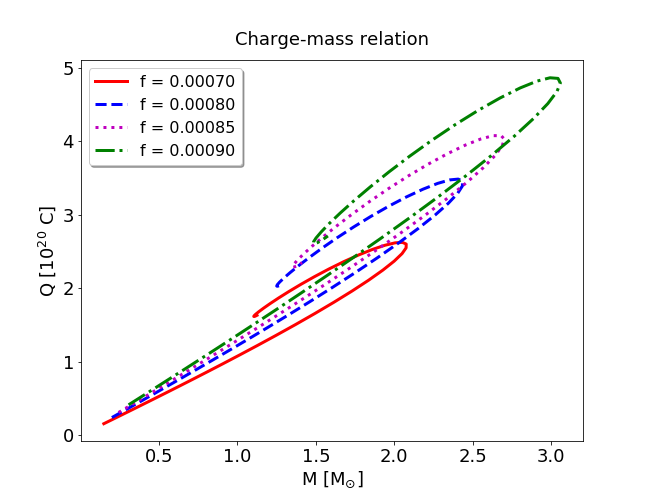}
\caption{Charge-mass relation.}
\label{fig:QM}
\end{figure}

We then build a sequence of 50 neutron stars for different values of the charge fraction, $f$. We present the mass-radius (Figure \ref{fig:MR}), the mass-central density (Figure \ref{fig:MRho}), and the charge-mass (Figure \ref{fig:QM}) relations for charged neutron stars with $f=(0.00070, 0.00080, 0.00085, 0.00090)$ to understand how the charge changes the star's overall structure. Notice that these values for {\it f} cover neutron stars with masses up to $3~M_{\odot}$ and still respect the limit $Q<M$ \citep{Arbanil_2013}.

To proceed with the falling times' calculations of the charges onto the neutron stars, we used the sequences obtained from $f=0.00080$, $f=0.00085$, and $0.00090$ - the first support maximum mass of $M\simeq 2.4~M_{\odot}$ and the latter of $M\simeq 3.0~M_{\odot}$ in conformity with recent detections \citep{Linares_2018, 2020_AbbottR}.

It is important to remark that the stellar models we calculated do not consider built-in rotations since the star's overall structure is practically not affected by the rotation in the slow-rotation regime assumed here consistent with the rotational period observed for all magnetars.

\subsection{The initial state of the system: the plasma horizon}

Assuming that the electric field's quadrupole moment dominates at a large radius, a ring of opposite charges will form at a distance of the star delimited by the region $r_{m}<r<r_{p}$, where $r_{p}$ is the so-called plasma horizon.

The plasma equilibrium condition 

\begin{equation}
    \Bigg[\frac{\Delta}{r^{4}}\Bigg]^{1/2}r^{3}\sin{\theta}\mid_{r=r_{p}}=\frac{Q}{B},
\label{plasmaHorizon}    
\end{equation} gives the plasma horizon, where $B = \frac{2~|\vec{\mu}|}{r^{3}}$, where $\vec{\mu}$ is the dipole moment of the star. $|\vec{\mu}|\sim 10^{30}~\textrm{emu}$ are the values of interest in our case.

The existence of the dead field lines avoids the neutron star spontaneous discharge. This condition should be metastable, but it should live long enough to produce a FRB.

Once we place the current ring at some radius (e.g., $r=25~\mathrm{M}$ in the right-hand side of equation \ref{plasmaHorizon}), we can calculate the plasma horizon for the case of the charged neutron stars.

\section{Results}
\label{results}

\subsection{Falling times}

As mentioned earlier, it is the charged neutron star's sudden discharge that triggers the FRB phenomenon. For this mechanism to work, the discharge must occur at a timescale short enough to produce a shock wave that, in turn, is the cause of the radio burst.

To test this hypothesis, for each of our 50 charged neutron stars, we first calculated the circular orbit radius, $r_{m}$, from $dV/dr = 0$, and then chose the falling radius, $r_{o}=0.95~r_{m}$. With $r_{o}$, we calculated the minimum energy of the orbit, $E_{min}$. Then, we set the energy of the falling particle to be $E=1.015~E_{min}$. This procedure is mostly similar to the one in \cite{liuT01}.

\begin{figure*}
  \centering
  
  \includegraphics[scale=0.35]{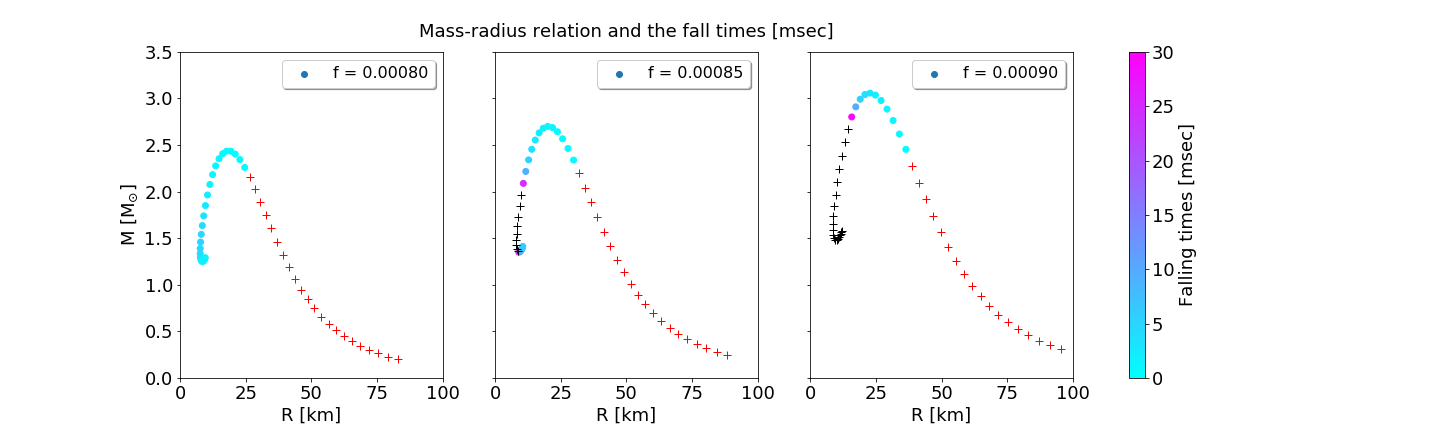}

  \caption{The falling times in milliseconds (see the color bar at right) of charged test particles onto the oppositely charged neutron stars in neutron star sequences built with the charge fraction of $f=0.00080$, $f=0.00085$ and $f=0.00090$ presented in the mass-radius diagram. We assumed a spin rate of 1 Hz for all stars. The red and black plus markers represent those stars in the sequence that there is no physical solution for the falling times, i.e., mostly because or $r_{m} < R$ or $M^{2} < Q^{2}$. Besides, the red plus markers are the stars in the stable branch of the sequence, and the black plus markers are the stars in the unstable one. The falling times range from 0.2 msec to 5 msec. We clearly see that the higher the charge fraction, the higher is the falling times.}
  \label{fallingTimes}
  
\end{figure*}

With all these quantities, we could evaluate the integral $\delta t$ (Eq. \ref{deltaT}) for the falling time. We present the results of this calculus in Figure \ref{fallingTimes}.

In Figure \ref{fallingTimes}, we present the falling times of charged test particles onto the oppositely charged neutron stars in the color bar at right. The bluer colors indicate shorter falling times while the pinkish colors indicate higher falling times. The red and black plus markers represent those stars in the sequence that there is no physical solution for the falling times, i.e., mostly because or $r_{m} < R$ or $M^{2} < Q^{2}$. Besides, the red plus markers are the stars in the stable branch of the sequence, and the black plus markers are the stars in the unstable one.

We built these sequences using charge fractions of $f=0.00080$, $f=0.00085$ and $f=0.00090$. The effect of the increase in the charge fraction is to increase the falling times. 

The vital information in Figure \ref{fallingTimes} is that the falling times are of the order of a few milliseconds, ranging mostly from $\simeq 0.2$ msec to $\leq 10$ msec, which makes the discharge at all sudden.

Besides, the smallness of the FRBs' timescales implies that the emission region is very compact, and the coincidence of this variability timescale with the falling timescales sets the whole emission scenario very close to the compact object. Below, we discuss the proper mechanism.

\subsection{Radiative processes}

The sudden discharge causes the magnetic field lines to disconnect and break. Immediately after, the field lines tend to reconnect, and this process of disconnection and then reconnection results in large electric currents parallel to the magnetic field lines while the strong magnetic shock wave generated in the process moves at the velocity of light throughout the plasma ahead. These electric fields accelerate particles at relativistic speeds, and curvature radiation is emitted. This mechanism is analogous to the one hypothesized by \cite{falcke01}, as also discussed by \cite{liuT01}.

Enclosed by the magnetosphere, the plasma ahead co-rotates with the neutron star due to its strong magnetic field as a rigid body, i.e., the plasma's velocity increases with the distance from the neutron star. The point where the velocity of plasma equals that of the light is the light cylinder, and it is the distance that the magnetic shock wave passes through, exciting the plasma collectively. That is the emission region.

The magnetospheric distance is given by

\begin{equation}
R_{mag} = \frac{c}{\Omega},
\end{equation} where $c$ is the velocity of light and $\Omega = 2\pi\nu_{s}$ is the angular velocity of the neutron star, $\nu_{s}$ being its spin rate. The corresponding travel time is given by

\begin{equation}
\tau = \frac{R_{mag}}{c}=\frac{1}{2\pi\nu_{s}},
\end{equation} obviously.

The curvature radiation emitted frequency and the correspondent emitted (incoherent) power per charge are given by

\begin{equation}
\nu_{e}=\frac{3}{2}\gamma^{3}\nu_{s}, \hspace{0.5cm} P_{e} = \frac{2\gamma^{4}e_{c}^{2}c}{3R_{mag}^{2}},
\end{equation} where $R_{mag}$ is the curvature radius of the emission, $e_{c}$ is the electron charge ($e_{c}=4.8\times 10^{-10}$ esu), and $\gamma$ is the Lorentz factor.

Now, FRBs' typical duration lies between one and ten milliseconds; see Figure 13.d in \cite{petroff02}.

\begin{figure}
  \centering
  
  \includegraphics[scale=0.35]{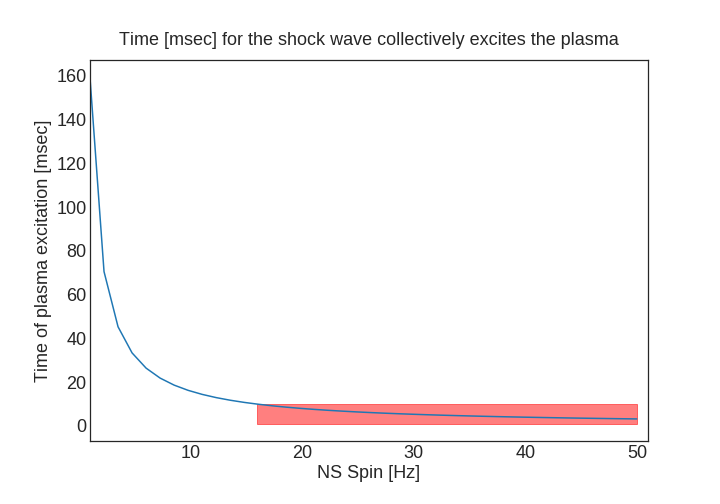}

  \caption{
Excitation times for the magnetic shock wave to excite the plasma collectively, $\tau=1/(2\pi\nu_{s})$. The reddish rectangle delimits the region for which the neutron star spin provides excitation times between one and ten milliseconds, the intervals where we find the bulk of the duration times of FRBs.}
  \label{excitationTimes}
  
\end{figure}

Thus, if $\tau$ lies between one and ten milliseconds, it is possible to relate the scenario to the FRB phenomena. In Figure \ref{excitationTimes}, we plot $\tau$ versus the spin rate of the neutron stars. The reddish rectangle shows the spin range that allows $\tau$ to be between one and ten milliseconds.

From Figure \ref{excitationTimes}, we readily see that the allowed values for the neutron stars' spin lie roughly between 16 and 50 Hz, or rotation periods between 0.0625 and 0.02 seconds, slow enough to keep very approximately the spherical symmetry of the star and well below the mass shedding limit, which characterizes the slow-rotation regime.

\begin{figure}
  \centering
  
  \includegraphics[scale=0.375]{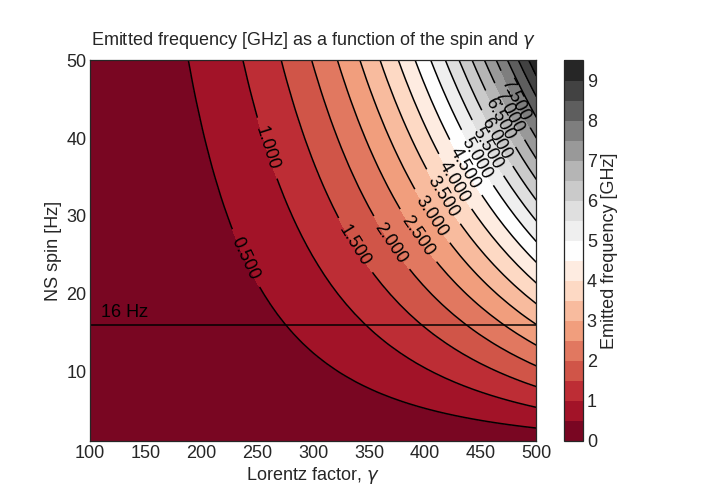}

  \caption{In the colored bar at right, we read the emitted frequency in units of GHz. For a given spin frequency, we learn J's allowed values that allow a given observed FRB frequency. The horizontal black line marks the minimum neutron star spin that allows for a millisecond duration of the radio burst.}
  \label{emittedFrequency}
  
\end{figure}

Rotation rates established we must now look for $\gamma$ values that provide emitted frequencies in the range we observe the FRBs. We show these values in Figure \ref{emittedFrequency}. FRBs are detected mostly at frequencies between 0.4 and 8 GHz \citep{petroff02}.

From Figure \ref{emittedFrequency}, we learn the pair of values [$\nu_{s}$, $\gamma$] that match the observed FRB frequency.

\begin{figure}
  \centering
  
  \includegraphics[scale=0.325]{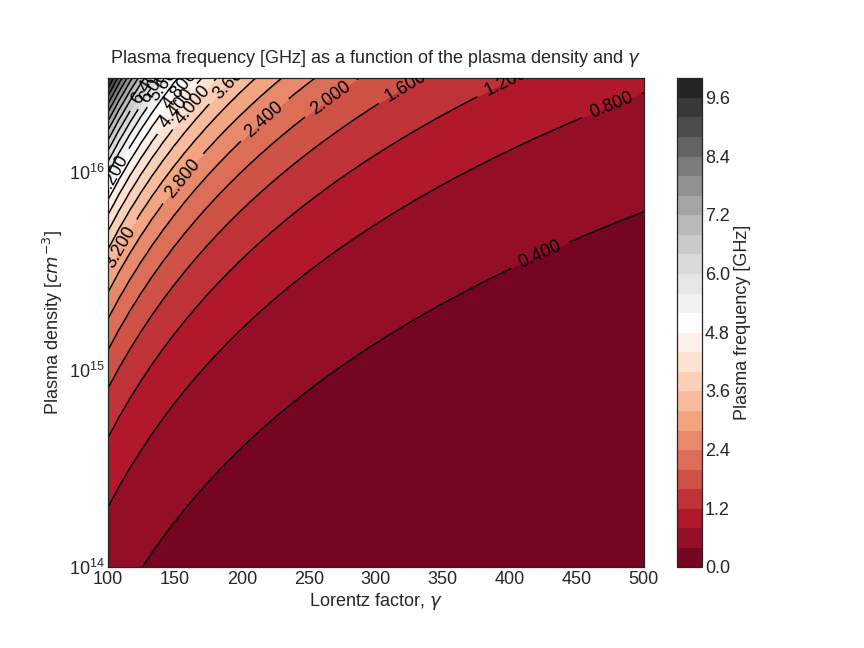}

  \caption{In the colored bar at right, we show the plasma frequency (in GHz units) as a function of the Lorentz factor and the plasma density.}
  \label{plasmaFrequency}
  
\end{figure}

Furthermore, for the radio burst to be observable, its frequency must be greater than the plasma frequency. For frequencies below the plasma frequency, the plasma is opaque to radiation. According to expression \ref{plasmaFrequencyExpression} below, the plasma frequency depends upon the Lorentz factor, $\gamma$, and the plasma density. We show the constraints in Figure \ref{plasmaFrequency}.

\begin{equation}
\label{plasmaFrequencyExpression}
    \nu_{plf} = \gamma^{-3/2}\Big(\frac{4\pi n_{e}e_{c}^{2}}{m_{e}}\Big)^{1/2},
\end{equation} where $n_{e}$ is the plasma density, and $m_{e}$ is the electron mass.

In the colored bar at the right of Figure \ref{plasmaFrequency}, we show the plasma frequency (in GHz units) as a function of the Lorentz factor and the plasma density.

Summarizing, from the observed FRB frequency, we learn the pairs ($\nu_{s}$, $\gamma$) corresponding to this frequency (See Figures \ref{excitationTimes} and \ref{emittedFrequency}). Then, from Figure \ref{plasmaFrequency}, we can select the proper pair ($\nu_{s}$, $\gamma$) by inspection, which plasma density provides the plasma frequency that allows the observed FRB frequency.

With the plasma density in our hands, we can calculate the total emitted power of a coherent pulse via $P_{tot}\sim (n_{e}V)^{2}P_{e}$ once we know the coherence volume, V, occupied by the plasma or, the other way around, to find the plasma coherence volume once we measure the total emitted power. Typically, $P_{tot}\sim 10^{42-43} erg/s$. See \cite{falcke01}, and \cite{petroff02}, and \cite{ghisellini01}, for example.

We stress that the typical values obtained for $\gamma$ and $n_{e}$ with the model proposed here are in line with, for example, \cite{ghisellini01}.

 \section{Discussion}
 \label{discussion}
 
\cite{ghisellini01} argue that a significant part of the FRBs emission arises through curvature radiation. We suggest the same radiative process due to the shock wave generated by the electric-magnetic field reconnection after the discharge. 

According to \cite{ghisellini01}, a challenge to the mechanism would be the cooling rate of the electrons, which implies a minimal path for the electrons to travel before the emission disappears. In other words, because the energetic loss is fast, the charges’ time travel must be small. Thus, they suggest two possible scenarios to circumvent the cooling problem. The first is many bunches of few particles, and the second is a re-acceleration mechanism based on strong electrical fields parallel to the magnetic field.
 
In the magnetar scenario we assumed that the re-acceleration mechanism comes into play to circumvent the cooling problem of the charged particles. Thus, the shock wave works as a re-acceleration source, and it takes only a few sub-seconds. The emission stops as soon as the shock dissipates because the radiative cooling time is short (10-12 sec). The reconnection processes end, and the system stabilize in a new configuration.

Together with the mechanisms discussed in \cite{ghisellini01}, the scenario we proposed also explains why FRBs are not detected in other wavelengths. \cite{ghisellini01} demonstrated that suppressing the inverse Compton scattering prevents the FRB from emitting in other wavelengths. In the model, the suppression occurs because of the symmetry in the scattered radiation in the comoving frame, which makes that the radiation bulk to happen along the electron velocity vector on the observation frame, reducing largely the scattered photons.

It is interesting to remark that \cite{falcke01} interpreted the FRB as the collapse of a supermassive neutron star into a black hole, triggering a mechanism very similar to the one discussed here and in \cite{liuT01}. In this sense, our model encompasses their model since the sudden discharge of neutron stars whose mass is very close to the maximum mass of the sequence can be just enough to cause its collapse. If the system survives, in other words, if the magnetar does not collapse after the discharge, the existence of repeater sources can be related to the time needed to magnetosphere charge again to produce a new discharge. We can suppose that this time can vary significantly depending on each system’s conditions, and one can not rule out the existence of a blast repetition after a long time. Maybe there are repeater sources classified as a single event because not enough time has passed for a new event.

Thus, constraining our model by recent observations (frequencies, durations, energetics; see \cite{petroff01}), we find that the FRB originates in a charged neutron star spinning at a rate varying from 16 to 50 Hz in an environment where the particles in the plasma trapped by the magnetic field of the star have Lorentz factors from 200 to 500. Besides, in our model, the surface magnetic field varies from $10^{9}$ G to $10^{13}$ G with the values for $|\vec{\mu}|$ we used but can be higher for other suitable choices of $|\vec{\mu}|$ without losing generality.

Regarding magnetars' masses that produce falling times compatible with the FRB timescales, we found an allowable mass range from 1.5 to 3 $\mathrm{M_{\odot}}$ (see Figure \ref{fallingTimes}), depending on the charge fraction f:

\begin{itemize}
\item $f=0.00080$ produces masses from 1.5 to 2.5, with most of its stars on the unstable branch;

\item $f=0.00085$ produces masses from 2.0 to 2.5, with stars equally distributed on the stable and unstable branches;

\item $f=0.00090$ produces masses from 2.4 to 3.1, with most of its stars on the stable branch.
\end{itemize}

The interesting point is that, in our model, we can relate the unstable branch to the non-repeaters since the discharge event could potentially collapse the system to a black hole and the stable branch to the repeaters since there would be time to recharge before a new sudden discharge, i.e., the discharge would not lead the system to a collapse. 

Hence, we have just seen how the millisecond falling timescale of charges onto an oppositely charged neutron star triggers a magnetic shock wave. This wave traveling at the velocity of light through a plasma medium in a timescale also of milliseconds, produces enough curvature radiation to feed an FRB.

The CHIME and STARE2 radio telescopes recently detected spacially and temporal coincident radio bursts during the G-Burst of the SGR 1935+2154 detected by INTEGRAL between April 28 and May 3 of 2020 \citep{mereghetti01}. This event was the first one where a soft gamma burst occurred almost concomitantly with a radio counterpart, and the detection strongly supports the FRB-SGR/Magnetars link \citep{mereghetti01}. The main properties of SGR 1935+2154 are period P = 3.25 seconds (spin around 0.31 Hz) and an inferred magnetic field of $2.2 \times 10^{14}$ G. Also, its energetics place it nearby other FRBs. Following our methodology, the plasma excitation time for this slow rotator is around 520 ms and requires a gamma factor between 1000 and 1200 for the radio band stays between 0.4 and 8 GHz.

Furthermore, the link FRB-magnetar allows explaining the periodicity in the sources better, indicating that the charging process in the short-time repeater sources is related to the spinning of the magnetars, or at least with some gravitational constraint. However, it does not mean that there are different processes present or absent in other FRBs sources so that we can have a variety of time scales of repeaters. Sometimes periodic, other times not. Detection of new periodic events and understanding how often they are will add light to the many possibilities of charging.

There are two other points that we must address regarding the model we proposed in this work. The first one concerns the possibility of constraining the equation of state of the neutron star or magnetar. Regarding this issue, our results show (see Figure \ref{fallingTimes}) that to have falling times in the acceptable range of FRB phenomena, the neutron stars must be massive and very compact. We also see that the smaller the {\it f} fraction, the lighter are the neutron stars, the smaller is the falling times, and most of the allowable stars are in the unstable branch (see Figure \ref{fallingTimes} and compare the leftmost panel with the rightmost one). The analysis we did about the {\it f} fraction led us to restrict its value in the range of 0.00070 to 0.00090, implying maximum masses for the neutron stars from 2.5 to 3.1, consistent with recent findings \citep{Linares_2018, 2020_AbbottR}. 

Thus, the quest for an equation of state that accommodates these high values for the maximum mass must take into account the effect of the charge on masses which is to increase the mass in comparison to the same equation of state without charge. In principle, (realistic) equations of state ruled out by recent observations can come into play again because of the effect of the charge.Thus, from our FRB model, it seems to be challenging to obtain some constraints on the equation of state of neutron stars, although we see a lack of investigations concerning the use of realistic nuclear matter equations of state in charged neutron stars in the literature \citep{2007_AlloyM, 2011_RotondoM}.

Compact remnants of compact objects mergers have opened new avenues for studying the equation of state of compact objects through gravitational wave detections resulting from the mergers themselves since many of the new findings suggest that at least one compact object should be an unusually high mass neutron star or a not yet proven low mass black hole. \citep[See][for example]{2010_OzelF,2017_AbbottB, 2018_LinaresM, 2019_ThompsonT, 2020_AbbottB}.


Such high mass neutron stars are significant to constrain the possible equations of state for these stars, at least in the context of non-charged compact objects. Notably, \cite{2020_AbbottR} reported the observation of a merger involving a 22.2-24.3 $\mathrm{M_{\odot}}$ black hole and a compact object with a mass of 2.50-2.67 $\mathrm{M_{\odot}}$. Thus, this compact object mass range becomes plausible if we consider the charge, as suggested by our work.

Very recently, using merger data from LIGO/VIRGO collaboration and timing data from the NICER satellite, \cite{2021_LiA} performed a Bayesian analysis of the maximum mass of neutron stars with a quark core.

\cite{2021_LiA} found that the most probable values of the hybrid star maximum mass are 2.36 $\mathrm{M_{\odot}}$ or 2.39 $\mathrm{M_{\odot}}$ depending on the equation of state used for the quark core, but with an absolute value of 2.85 $\mathrm{M_{\odot}}$ either way (see also \cite{2020_LiA} for a salutary discussion about kaons and hyperons on the equation of state of neutron stars and \cite{2003_MalheiroM} about strange quark matter). These results suggest that GW 170817 could be a massive rotating neutron star \citep{2020_AiS}. Notice that the values they found ($\sim$ 2.4 $\mathrm{M_{\odot}}$) are slightly below the values we found in our work so that a small charge fraction would suffice to bring these quarkionic equations of state into play.


The second point is regarding the event rate for FRB  in our theory. Firstly, it is essential to remember that our model does not necessarily request the magnetar collapse since the repeaters are explained precisely by the formation of new pulses in the original (non-cataclysmic) progenitor. Nevertheless, we can estimate the number of collapsing magnetars from a statistical neutron star mass distribution \citep{2021_RochaL}. From a simulation of 5000 distributions obtained from the posterior predictive check on two-Gaussian models without truncation, Figure 2 in \citet{2021_RochaL} shows a 14.4\% probability of finding neutron stars with masses $>2.43M_\odot$. These results could help us to estimate the fraction of collapsing magnetars related to the non-repeater FRBs in our model, but the events rate related to the repeaters is still missing.

The repeaters appear depending on the system recharge, but we cannot correctly predict the timescale of the recharge due to the many uncertainties involved in the process. However, if we cannot fix the event rate for all the FRBs, on the other hand, we can explain different observational rates because of the flexible pattern of our model.  In principle, it can vary by a significant amount between different sources. It is not a limitation of the model but rather one of its triumphs, as it can explain different event rates. Our model allows the existence of repeaters after many years from the first FRBs, in the case of a given source having a slow recharge process. If such a case is possible, even the actual total rate is imprecise because it does not have enough time to get a reasonable rate estimate in a yearly timescale. Notice that the time scale between consecutive FRBs can vary even in the same source, explaining actual repeaters with no apparent periodicity.


Finally, an estimation of the number of magnetars in the Galaxy ($\geq 10^{7}$) \citep{2015_TurollaR, 2014b_ReaN,2010_KaspiV} could give us further hints in the direction of the event rate of our FRB model.

\section{Conclusion}
\label{conclusions}

We propose a new model for FRBs based on a sudden discharge of a young magnetar due to oppositely charged particles falling onto the star in a timescale to reach the target short enough to trigger a shock wave that excites the plasma, producing the burst in a milliseconds timescale.

Our modeling begins with the generalized Tolman-Oppenheimer-Volkoff equations for a spherical, massive, and charged body, which are solved for the interior spacetime and matched to the exterior one using the Reissner-Nordstr\"om metric. This metric describes the spacetime outside the charged magnetar well in the static approximation since magnetars have very long periods in the neutron stars period scale. Besides, to solve the system of equations, we assumed that the charge distribution is proportional to the mass-energy density, whose proportionality constant is the charge fraction {\it f}.

Once we solved the system, we calculated the circular orbits from the effective potential and chose the orbit the particles would fall onto the magnetar, discharging it. With this, we demonstrated that the falling times of the opposed charged particles onto the charged magnetar are short enough (5 to 30 msec, see Figure \ref{fallingTimes}) to trigger a shockwave that, in turn, collectively will excite the plasma around.

The next step was to discriminate the values the magnetar spin should assume (15 to 50 Hz, see Figure \ref{excitationTimes}) for the excitation times to be of the proper order of magnitude to match the FRBs durations (5 to 20 msec, according to the most up-to-date literature). The spin scale so found justifies and corroborates to a certain degree the magnetar scenario.

Thus, knowing the typical emitted frequencies of FRBs (0.4 to 8 GHz) and the proper neutron star spins, we could find the environmental conditions, i.e., the Lorentz factor values range of the plasma that best describes that emitted frequencies (see Figure \ref{emittedFrequency}). Then we used the plasma frequency to constrain the whole scenario since the plasma can not be opaque to the emitted radiation burst (Figure \ref{plasmaFrequency}).

All the calculations mentioned above lead to values that fit well the magnetar scenario for the FRB phenomena, i.e., magnetar spin rate in the range 16 to 50 Hz, Lorentz factor in the range 200 to 500,
the magnetic field from $10^{9}$ to $10^{13}$ G, emitted power of $10^{42}$ to $10^{43}$ erg/s.

Thereby, the scenario we propose is suitable for explaining the observables of the FRB phenomena. However, to fit best the observables, the preferred magnetar's masses (and their radii) must lie close to the maximum mass of the sequence of stars allowed by the equation of state, either on the stable or the unstable branches. For f = 0.00085, for example, M $\simeq$ 2.5 $\mathrm{M_{\odot}}$ and R = [12 - 30] km. In turn, to be in the stable or unstable branches opens an avenue for explaining the existence of repeaters and non-repeaters. When the magnetar belongs to the stable branch, the discharge may not be enough to collapse the magnetar, which then lives for another round of charging and discharging the magnetosphere. In the other case, the discharge would collapse the magnetar to a black hole, producing a one-time event.

Finally, the discharge of a magnetar is expected to be a big energetic event capable of producing a powerful shock wave that produces a bright and short-lived radio emission. Our findings are in line with recent detections and theoretical advances for explaining FRBs, and we expect that new detections by the CHIME Collaboration may confirm the magnetar scenario and the existence of very massive charged neutron stars, even if short-lived due to their most expected collapse.

\section*{Acknowledgements}

MGBA acknowledges CNPq project 150999/2018-6. PHRSM thanks CAPES for financial support. MM is grateful to CAPES and CNPq financial support. The authors acknowledge the FAPESP Thematic Project 2013/26258-4.

\section*{Data Availability}


No new astrophysical or astronomical data were generated or analyzed in support of this research. All data in this paper are computer-generated through owned software. The computational data underlying this article will be shared on reasonable request to the corresponding author.

\clearpage



\bibliographystyle{mnras}
\bibliography{export-bibtex} 








\bsp	
\label{lastpage}
\end{document}